\documentclass{emulateapj}
\usepackage{apjfonts}

\slugcomment{To appear in the  ApJS Spitzer Issue}

\shorttitle{IRAC Observations of PN}
\shortauthors{Hora et al.}

\begin{document}


\title{IRAC Observations of Planetary Nebulae
    }


\author{Joseph L. Hora\altaffilmark{1}, William B. Latter\altaffilmark{2},
Lori E. Allen\altaffilmark{1}, Massimo Marengo\altaffilmark{1},
Lynne K. Deutsch\altaffilmark{1,3}}
\author{Judith L. Pipher\altaffilmark{4}} 

\altaffiltext{1}{Harvard-Smithsonian Center for Astrophysics, 60 Garden St., 
MS-65, Cambridge, MA  02138-1516}
\altaffiltext{2}{Spitzer Science Center, MS 220-6, Caltech, Pasadena, CA 91125}
\altaffiltext{3}{Deceased (2004/04/02).  Lynne will be deeply missed by her
friends and colleagues on the IRAC team.}
\altaffiltext{4}{University of Rochester, Dept. of Physics and Astronomy, Rochester, NY
14627-0171}



\begin{abstract}
We present the initial results from the Infrared Array Camera (IRAC) 
imaging survey of planetary nebulae (PN). 
The IRAC colors of PN are red, especially when considering the 8.0 $\mu$m band.
Emission in this band 
is likely due to contributions from two strong H$_2$ lines and a
[\ion{Ar}{3}] line in that bandpass.  IRAC is sensitive to the emission in
the halos as well as in the ionized regions that are optically bright.
In NGC 246, we have observed an unexpected
ring of emission in the 5.8 and 8.0 $\mu$m IRAC bands not seen previously
at other wavelengths.  In NGC 650 and NGC 3132, the 8.0 $\mu$m emission is
at larger distances from the central star compared to the optical and other
IRAC bands, possibly related to the H$_2$ emission in that band and the
tendency for the molecular material to exist outside of the ionized zones.
In the flocculi of the outer halo of NGC 6543, however, this trend is
reversed, with the 8.0 $\mu$m emission bright on the inner edges of the
structures.  This may be related to the emission mechanism, where the H$_2$
is possibly excited in shocks in the NGC 6543 halo, whereas the emission
is likely fluorescently excited in the UV fields near the central star.
\end{abstract}

\email{jhora@cfa.harvard.edu}





\section{Introduction}

In recent years the study of stellar ejecta released during final
evolutionary stages has provided many new insights
into stellar evolution. One reason for these major leaps forward have
come from the advancement of near-IR imaging and spectroscopy, as
well as space-based platforms (HST, ISO) providing new views of the
stellar death process. 
 Mass lost from stars on the Asymptotic Giant Branch (AGB) is
observable as extended envelopes around planetary nebulae (PN). 
A significant amount of material released to the ISM by 
PN and supernovae is in the form of silicate and carbonaceous
dust. Tiny carbon grains, or polycyclic aromatic hydrocarbons (PAHs), have
strong emission features throughout much of the near- and mid-IR
spectral regions -- in addition to numerous atomic and
molecular features from ionized and shock heated gas. 

Several of the brightest PN have been imaged 
in the mid-IR and near-IR at high spatial resolution from the ground
\citep{hora90, hora93,latter95},
but the sensitivity of the Infrared Array Camera
\citep[IRAC;][]{fazio04} and the other instruments on the Spitzer Space Telescope
\citep{werner04} will allow us to expand our knowledge beyond 
what was possible on the ground or previous space missions.
ISO  provided an initial high-sensitivity look at the 
characteristics of PN in the mid- and far-IR 
\citep{cox98,hora99,fong01,castro01}.  One large nearby
PN that was observed
by Cox et al. was NGC 7293 (The Helix). The mid-IR
emission is dominated by emission from molecular hydrogen in the ground
rotational state.  The brightest line, the $\nu=$0--0 $S(5)$ at 6.91 $\mu$m
and the $S(4)$ line at 8.02 $\mu$m fall within the 
IRAC 8.0 $\mu$m  band, so
that band is a good tracer of the H$_2$ in PN.  There will also be H$_2$ 
emission in the other IRAC bands, e.g. the  $\nu=$0--0 $S(7)$ line
at 5.51 $\mu$m in the 5.8 $\mu$m band, and the $S(9)$ line at 4.69 $\mu$m in the 4.5 $\mu$m band, and a large number of transitions in the 3.6 $\mu$m band \citep{black87}.  IRAC also will detect emission from
atomic lines present in the bands, 
e.g. the Br$\alpha$ H line at 4.05 $\mu$m in the 4.5
$\mu$m band, and the 6.98 $\mu$m [\ion{Ar}{2}] and 8.99 $\mu$m 
[\ion{Ar}{3}] line in the IRAC 8.0 $\mu$m band.  If emission from 
PAHs at 3.3, 6.2, and 7.7 $\mu$m are present,
they will contribute to the emission measured in the 
3.6, 5.8, and 8.0 $\mu$m IRAC bands.

With IRAC,
we are conducting a program to
observe 35 PN (program ID 68).  A related project with the Multiband 
Imaging Photometer for 
Spitzer (MIPS)   and Infrared Spectrograph (IRS) instruments 
led by Latter (program ID 77)
will obtain  24-160 $\mu$m images and 5 -- 40 $\mu$m
spectra of PN. 
In this paper we present images of six of the PNe in the 
IRAC program, which were the first six objects observed since
launch.  The PN are Hubble 12 (Hb 12), 
NGC 246,  NGC 650,
NGC 2440,
NGC 3132,
 and NGC 6543.

\section{Observations and Data Reduction}


The IRAC observations were obtained 
using the 30-sec ``High
Dynamic Range'' (HDR) mode.  This mode  takes pairs of images with 
1.2 and 30 sec frame times at each position in all four channels 
(3.6, 4.5, 5.8 and 8.0 $\mu$m; see \citep[see][for definitions of the bands]{fazio04}.  Between 5 and 18 dither positions
were obtained on each source. 
For objects that did not approach saturation, the 30 sec frame time 
data were used.  For the objects that were close to or above saturation in 30 sec, the 1.2 sec frames were mosaiced and used to fill in the brightest regions 
of the images.  

The Basic Calibrated Data (BCD) products from the Spitzer Science Center (SSC) 
pipeline  were used to construct the mosaic images for all objects. The BCD products have the
instrumental signatures removed from the data and are calibrated in units of MJy/sr, based on the 
calibration derived during the first few months of the mission from 
measurements of standard stars \citep{fazio04, horas04}.  Two further reduction steps
were performed on the BCD -  first, a sky background frame was constructed from
the median of the off-source images, and
subtracted from the BCD frames.  Also, some 
artifacts in the pipeline images caused by bright 
stars were removed  by forcing the column or row median in regions with no sources to be equal to that of
adjacent columns or rows.  The individual BCD images were combined into
a single image for each channel and frame time using the SSC mosaicer.
The final images have a linear pixel size 1/4 that of the input pixels (1/16 of the area).
For flux calibration, the zero magnitude fluxes in the IRAC
bands were taken to be 277.5. 179.5, 116.6, and 63.1 Jy for channels 1-4, respectively.  A
correction for the extended emission was applied to the 
fluxes as described in the Spitzer Observer's 
Manual\footnote{http://ssc.spitzer.caltech.edu/documents/som/}.

\section{Results and Discussion}
The IRAC images of the nebulae are presented in Figures 1 and 2.  
Color images of five of the PN are shown in Figure 1.  The color images are composites of all four IRAC bands,
as described in the figure caption. Images of the individual bands are shown in Figure 2.


\begin{figure}
\figurenum{3}
\includegraphics[scale=0.43, angle=0]{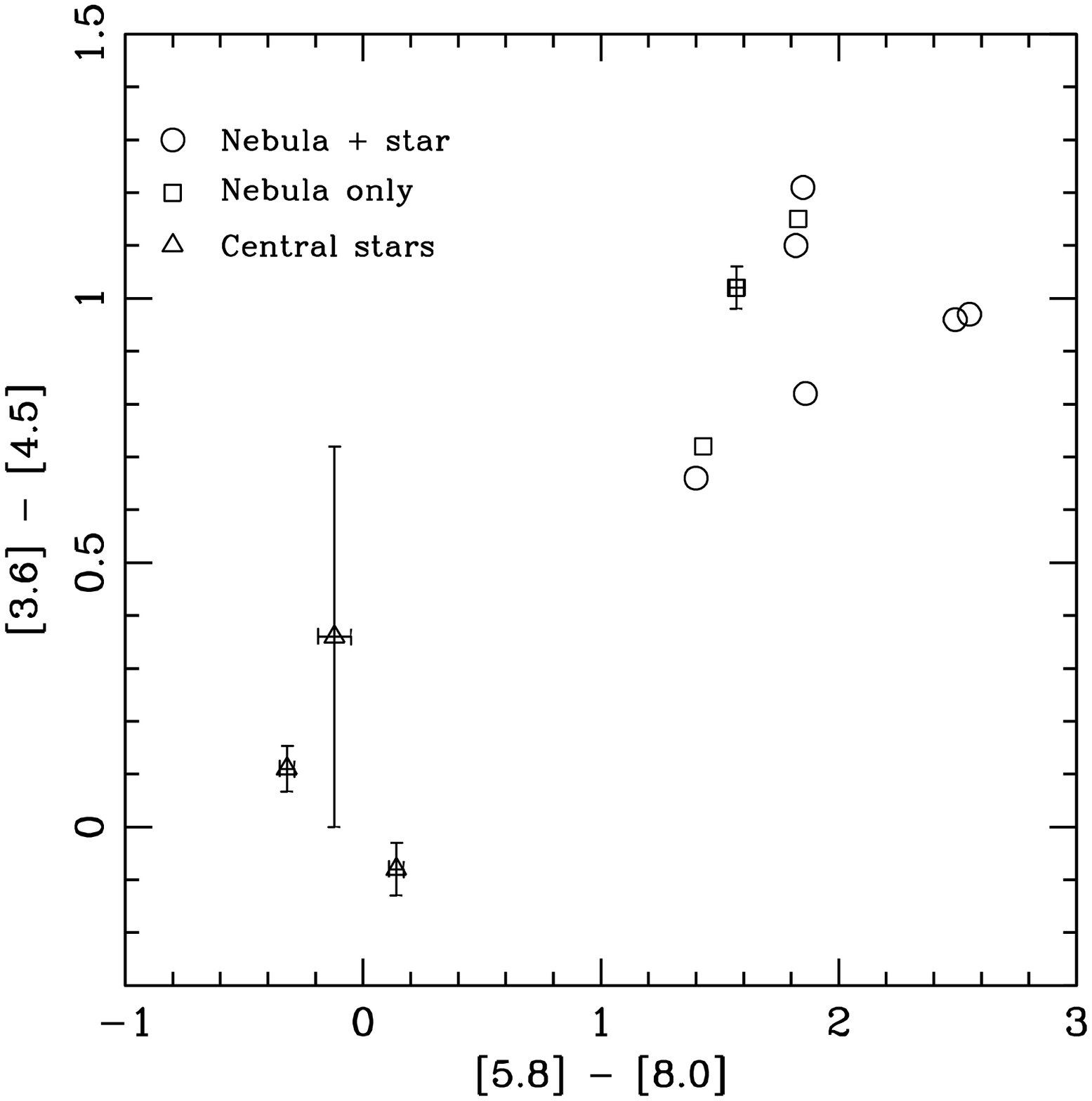}
\caption{Planetary nebulae colors from the IRAC data in Table~\ref{tbl-1}.
The total nebula+central star, nebula-alone, and central star-alone colors
are plotted, for those nebulae where it was possible to separate them out.  For NGC 6543, the
 color of the brightest knot in the extended halo is shown. The error bars for
the central stars are shown for each point, and a representative error bar
for the nebulae are shown for one point for clarity.}
\end{figure}

\begin{figure}
\figurenum{4}
\includegraphics[scale=0.34,angle=-90,clip=true]{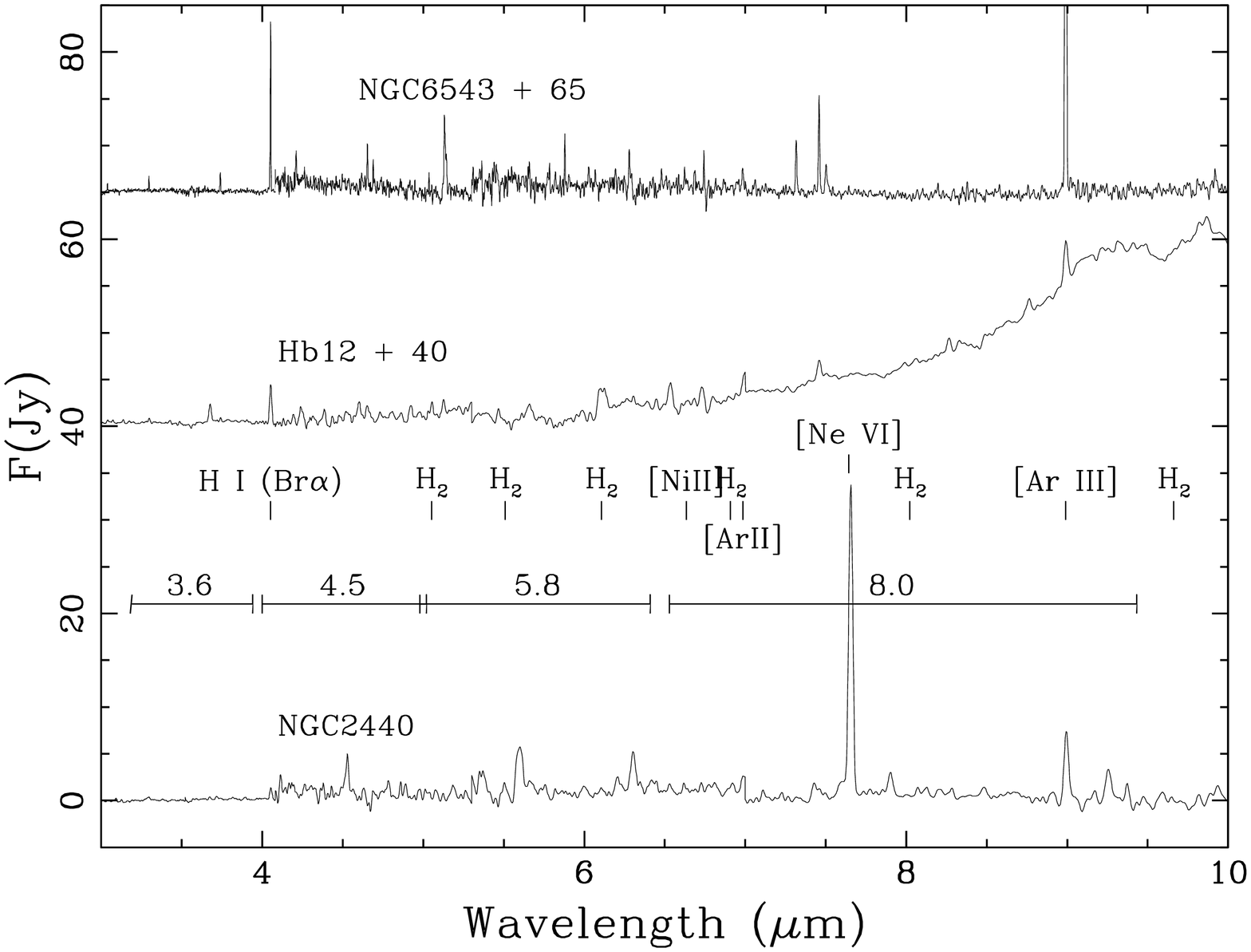}
\caption{
Spectra of Hb 12, NGC 2440, and NGC 6543, from ISO observations 43700330, 
72501762, and 02400714, respectively.  These are based on the on-line 
pipeline products downloaded from the ISO database.
The spectra of NGC 6543
and Hb 12 have been offset vertically by the indicated amounts.  Some
of the atomic and H$_2$ lines have been indicated. Also shown are the
wavelength ranges between the half-power points of the IRAC bands.  See
\citet{fazio04} for details on the IRAC spectral response function.
}
\end{figure}






\begin{deluxetable}{lcrrrr}
\tabletypesize{\scriptsize}
\tablecaption{Planetary Nebulae Magnitudes \label{tbl-1}}
\tablewidth{0pt}
\tablehead{
 & \colhead{Aperture}\\
\colhead{Source} &\colhead{(arcsec)} & \colhead{[3.6]} & \colhead{[4.5]} & \colhead{[5.8]} & \colhead{[8.0]} 
}
\startdata
Hb 12    &24.4\tablenotemark{a}&6.94&5.98&5.10  &  2.61\\
NGC 246& 305\tablenotemark{b}&8.91  &  7.80  &  7.59  &  5.77  \\
NGC 650&91.5\tablenotemark{b}&9.72  &  8.51  &  8.78  &  6.93\\
NGC 2440 &81\tablenotemark{b}&7.83  &  7.02  &  6.05  &  4.19\\
NGC 3132& 130\tablenotemark{b}&7.79  &  7.13  &  6.65  &  5.25\\
NGC 6543&48.8\tablenotemark{a}&7.56  &  6.59  &  6.87  &  4.32\\
NGC 246neb&305\tablenotemark{b}&8.97  &  7.82  &  7.60  &  5.77\\
NGC 3132neb&130\tablenotemark{b}&7.92  &  7.21  &  6.69  &  5.26\\
NGC 6543knot&92$\times$52\tablenotemark{c}& 11.11  &  10.09  &  9.33  &  7.75\\
NGC 246-cstar&3\tablenotemark{a}&12.02  &  12.10  &  12.01  &  11.86\\
NGC 650-cstar& 3\tablenotemark{a}&16.06  &  15.70  &  16.42  &  16.54\\
NGC 3132-cstar&3\tablenotemark{a}& 10.09  &  9.98  &  10.00  &  10.32\\
\enddata
\tablenotetext{a}{Circular aperture diameter}
\tablenotetext{b}{Square aperture width}
\tablenotetext{c}{Long dimension is horizontal direction in orientation shown in
Figure 1}
\end{deluxetable}

\subsection{IRAC colors of PN}
The Vega-relative magnitudes and colors of the PN presented in this paper are
shown in Table~\ref{tbl-1}.  The magnitudes for the nebula were estimated by 
summing all of the detectable emission from the nebula or the particular 
component, and subtracting the emission from the field stars in that part
of the image.  Column 2 indicates the size and shape of the area summed, centered on
the central star of the PN.  
In general the total field star flux was small compared to
the nebular flux, ranging from  $\sim$15\% in the 3.6 $\mu$m band for a couple of 
the PN to less than 3\% in the 8.0 $\mu$m band.  Where it was possible to 
extract the central star separately from the PN, those magnitudes and colors
are also presented, as well as the nebula without the central star.  Since the
nebulae dominate the flux from the object, the nebula-alone colors are almost
the same as the total object colors.

The IRAC color data are plotted in Figure 2.  The [3.6]-[4.5] color
is in the $\sim$0.6 -- 1.2 range for the sample, and the [5.8]-[8.0] color is
in the $\sim$1.4 -- 2.5 range.  A typical error bar for the nebula is shown for
one of the points; calibration uncertainties at this time early in the mission
are thought to be 
 0.1 mag or less in all channels.
Some of the [5.8]-[8.0] colors are redder than expected,
e.g., \citet{whitney03} predicted [5.8]-[8.0] PN colors of less than 2.
The red [5.8]-[8.0] color is possibly due to several factors, 
including continuum emission from warm dust (e.g., Hb 12), or line emission 
from forbidden atomic lines or H$_2$ lines in the broad 8 $\mu$m band (e.g., NGC 6543; see discussion
below).

The ISO-SWS \citep{kessler96, degraauw96} 
spectra of three of the PN are shown in Figure 4.  
The full spectrum of NGC 2440 was presented by \citet{salas02}.  
We show these spectra to illustrate representative spectra from some
of the PN in the mid-IR.  
Several atomic and H$_2$ line positions are shown, not necessarily
identifying them in the spectrum, but to indicate their position relative to
the IRAC bands.  No PAH emission is seen in the PN spectra shown in Figure 4. 
Also, the colors of the PN presented here are much redder than expected
if the emission was caused by only PAH lines \citep{li01}.

\subsection{Individual Nebulae}
\subsubsection{Hb 12}
Hb 12 is a young bipolar
PN with a complex inner structure inside the bipolar 
lobes, all of which is seen in fluorescently-excited
H$_2$ emission in the near-IR \citep{hora96}.
The IRAC image shows very little extended emission near the core,
so the images are not presented here.  The ISO spectrum in 
Figure 4 shows a rising continuum towards longer wavelengths,
which is consistent with the observed IRAC colors.
Lines of H$_2$ are also present, e.g. the 0 -- 0 $S(5)$ line
at 6.91 $\mu$m, and forbidden lines such as [\ion{Ar}{3}] at 8.99 $\mu$m.

\subsubsection{NGC 246}
NGC 246 is an oval-shaped, high-excitation PN approximately 
4.5 arcmin in diameter.  
 Recently, \citet{saint03} imaged NGC 246 in several near-UV lines 
and detected structure in the inner parts of the nebula.  In particular, the image in 
the [\ion{Ne}{5}] line at 342.9 nm shows bright arcs near the central star in an ``hourglass'' or bipolar structure near, but not centered on, the central star.  
The remarkable feature in the IRAC images is the oval structure that
appears brightest in the 5.8 and 8.0 $\mu$m bands.  This is possibly an inclined
ring, and is not centered on the central star or the outer shell of the
nebula.  The brightest part of the PN is the C-shaped feature to the SE of the central star, which is also brightest in the [\ion{Ne}{5}] image.
The cavity that forms the northern part of the hourglass is evident in the IRAC images, 
particularly the 4.5 $\mu$m band, but is not as distinct as in the UV image.

\subsubsection{NGC 650}
NGC 650 is a large, high-excitation PN with bipolar structure.  
The kinematic study of \citet{bryce96} showed that there are several components
including a central ring with two attached inner lobes, and outer lobes
with much lower expansion velocities, and a higher velocity polar cap on the 
SE side.  In the near-IR, the emission is dominated by H$_2$ that is brightest
in the central 2 arcmin region \citep{kastner96}.

All of the major components are seen in at least one of the IRAC 
bands.  The IRAC bands likely contain emission from H$_2$
as well as atomic lines.  The emission seen in the 8.0 $\mu$m band is more 
extended in the central ring and the lobes than the shorter bands, as evidenced by the reddish color on the outer edges seen
in Figure 1.  The 8.0 $\mu$m emission in the SW
extends towards the wall of one of the lobes, giving it the appearance of
a red arrow pointing to the SW. 
The H$_2$ emission in other PN, e.g. M 2-9 \citep{hora94}, or NGC 7027 \citep{latter95},
is predominantly  on the outer edges of the nebula, delineating the
ionization front.
\citet{bryce96} detected a polar cap in the SE part of the nebula with a higher
velocity than the surrounding shell.  The cap is faintly visible in the IRAC
bands, clearest in the 8.0 $\mu$m image.  We see a faint 
extended structure on the NW side that could be the opposing cap.
Also, $\sim$50\arcsec~
 to the SE of the central star there are three 
small emission features
about 5\arcsec~ long, oriented radially from the central star. They are bluer 
than the other features, and could be 
material ejected along the poles of the PN.

\subsubsection{NGC 2440}
The PN NGC 2440 is a high-excitation, point-symmetric nebula with at least three sets of opposing knots
of emission.  Around the inner parts of the nebula is a circular shell of emission, not centered on the symmetric inner structure.  There are radial ``spokes'' of emission 
extending from the central knots to the outer circular ring at several locations.  These 
structures are all seen in H$_2$ emission in the near-IR \citep{latter95, latter97}.
 The appearance 
in the IRAC bands is similar, with all the major features apparent. The relative brightness
of the features varies in the bands, with the outer shell much fainter at the longer
wavelengths.  The ISO spectrum doesn't show any bright emission lines, except possibly the 7.64 $\mu$m line of \ion{Ne}{6} which would be detected in the 8.0 $\mu$m band.  

\subsubsection{NGC 3132}
NGC 3132 appears ellipsoidal in low resolution images, however a simple
shell model does not explain all of the observed characteristics.  
\citet{monteiro00} propose a model in which the nebula has an hourglass
structure that is being viewed at 40\degr\ relative to the light of sight.  
This structure explains the details of the morphology and the low central 
density observed in the optical images.  

The IRAC images are similar to the optical and near-IR in the central regions,
with the 8.0 $\mu$m emission relatively brighter in the outer regions.
One interesting feature in the IRAC images is the faint extended emission 
approximately N-S along the major axis of the bright shell.  There are what
appears to be partial rings, and filaments connecting the rings radially 
with respect to the central star, and many knots and loops of emission that
extend up to approximately three times the size of the bright optical nebula. 
It is surprising given the proposed model that this extended emission would
appear along this axis, rather that the perpendicular axis where the major
axis of the lobes are proposed to exist. In fact, the nebula does not
appear to extend very much at all in the E-W direction beyond the bright
optical nebula.

\subsubsection{NGC 6543}
NGC 6543 (the ``Cat's eye'') is another point-symmetric nebula with a complex set of 
bubbles, rings, shock fronts, jets, and fast, low-ionization emission-line regions \citep[FLIERs; see][]{balick93,balick94}. 
These are most clearly visible in the HST images of this PN, e.g., in lines of 
[\ion{N}{2}] and [\ion{O}{3}] \citep{reed99}.  The near-IR appearance is similar
to the optical \citep{latter95}, with the prominent features being the oval-shaped rings at right angles to each other.  The IRAC images are similar to the near-IR and optical broad-band images, also showing the oval rings as the main features.  The FLIERs are more 
prominent at these wavelengths, especially in the 5.8 and 8.0 $\mu$m bands.  The ISO
spectrum of NGC 6543 shows bright lines of H (Br$\alpha$ at 4.05 $\mu$m) and [\ion{Ar}{3}]
at 8.99 $\mu$m which might affect the flux in the 4.5 and 8.0 $\mu$m bands, but otherwise the
emission is not detected in the 3 -- 10 $\mu$m range of the ISO spectrum in 
Figure 4.  

In addition to the bright inner region, there is a faint roughly circular halo extending
to a radius of approximately 170\arcsec~ \citep{balick92, balick01}.  The halo has a faint smooth
component, but near the nebula exhibits concentric rings 
in [\ion{O}{3}], and 
the outer rim is dominated by cometary-shaped ``flocculi'', the brightest 
being directly west
 of center. The appearance of the core and outer halo
is similar in the IRAC bands (the concentric rings are not observable at IRAC's
angular resolution).
There are some differences between the optical and IRAC images in the 
structure of the largest knot.
The knot is relatively brighter
at longer wavelengths, and the peak moves farther away
from the center  as one moves longer in wavelength.  In the 8.0 $\mu$m image,
the brightest spot appears behind the front of the cometary front, and there is much more
structure S of the knot.  Recent ground-based images \citep{hora04} show that the
knot has significant H$_2$ emission at 2.12 $\mu$m, so there is likely 
some emission from the 0 -- 0 $S(7)$ line in the 5.8 $\mu$m band and the $S(4)$ and $S(5)$
lines in the 8.0 $\mu$m band.

\section{Conclusions}
The IRAC colors of PN are red, especially when considering the 8.0 $\mu$m band
which is likely due to contributions from two strong H$_2$ lines and a 
[\ion{Ar}{3}] line in that bandpass, in addition to thermal continuum emission
from dust.  PAH emission is not detected in the objects presented here, but is expected
to be present in other PN. 
In NGC 246, we have observed an unexpected
ring of emission in the 5.8 and 8.0 $\mu$m IRAC bands not seen previously
at other wavelengths.  In NGC 650 and NGC 3132, the 8.0 $\mu$m emission is 
at larger distances from the central star compared to the optical and other
IRAC bands, possibly related to the H$_2$ emission in that band and the 
tendency for the molecular material to exist outside of the ionized zones.
In the flocculi of the outer halo of NGC 6543, however, 
the 8.0 $\mu$m emission is brightest on the inner edges of the 
structures.  This may be related to the emission mechanism, where the H$_2$
is possibly excited in shocks in the NGC 6543 halo, whereas the emission
is likely fluorescently excited in the UV fields near the central star.



\acknowledgments
This work is based on observations made with the Spitzer Space Telescope,
which is operated by JPL, Caltech,
under NASA contract 1407. Support for this work was provided
by NASA through Contract Number 1256790 issued by JPL/Caltech.
Support for the IRAC instrument was provided by NASA through Contract
Number 960541 issued by JPL.



Facilities: Spitzer/IRAC.

\clearpage


\begin{figure}
\figurenum{1}
\includegraphics[scale=1.7]{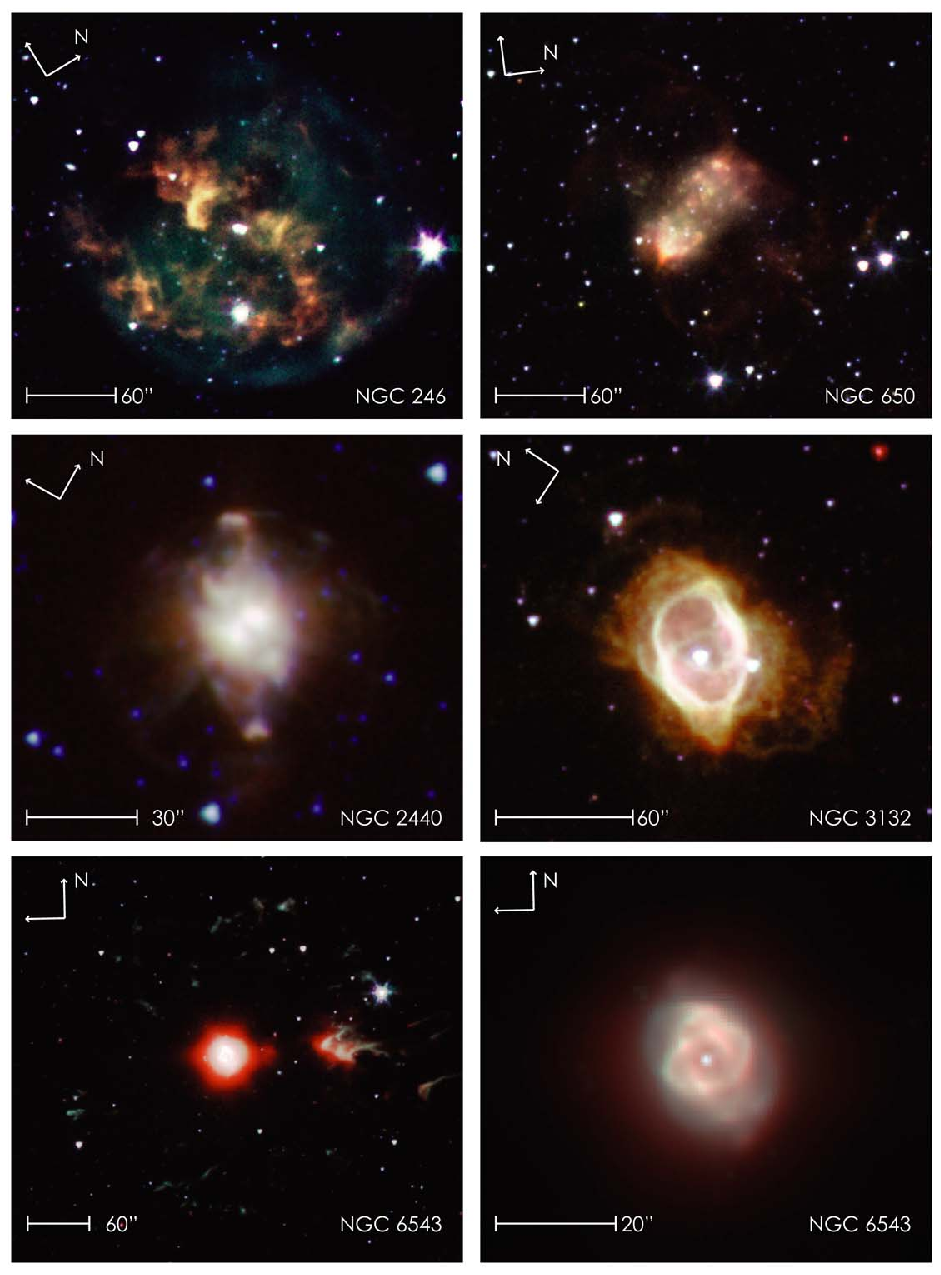}
\caption{(Plate 1:) Planetary nebulae color images.  The image orientations and scales
are shown on each image.  The images are displayed in a histogram equalized color scale (with the 
exception of the close-up of NGC 6543, which uses a square root color 
scale). The 4 IRAC bands have been reduced into the RGB color space 
using the following scheme:
- IRAC 8.0 $\mu$m = ``red'',
 IRAC 5.8 $\mu$m = ``cyan'',
 IRAC 4.5 $\mu$m = ``yellow-green'',
 IRAC 3.6 $\mu$m = ``purple''.
}
\end{figure}

\clearpage
\begin{figure}
\figurenum{2}
\includegraphics[scale=1.0]{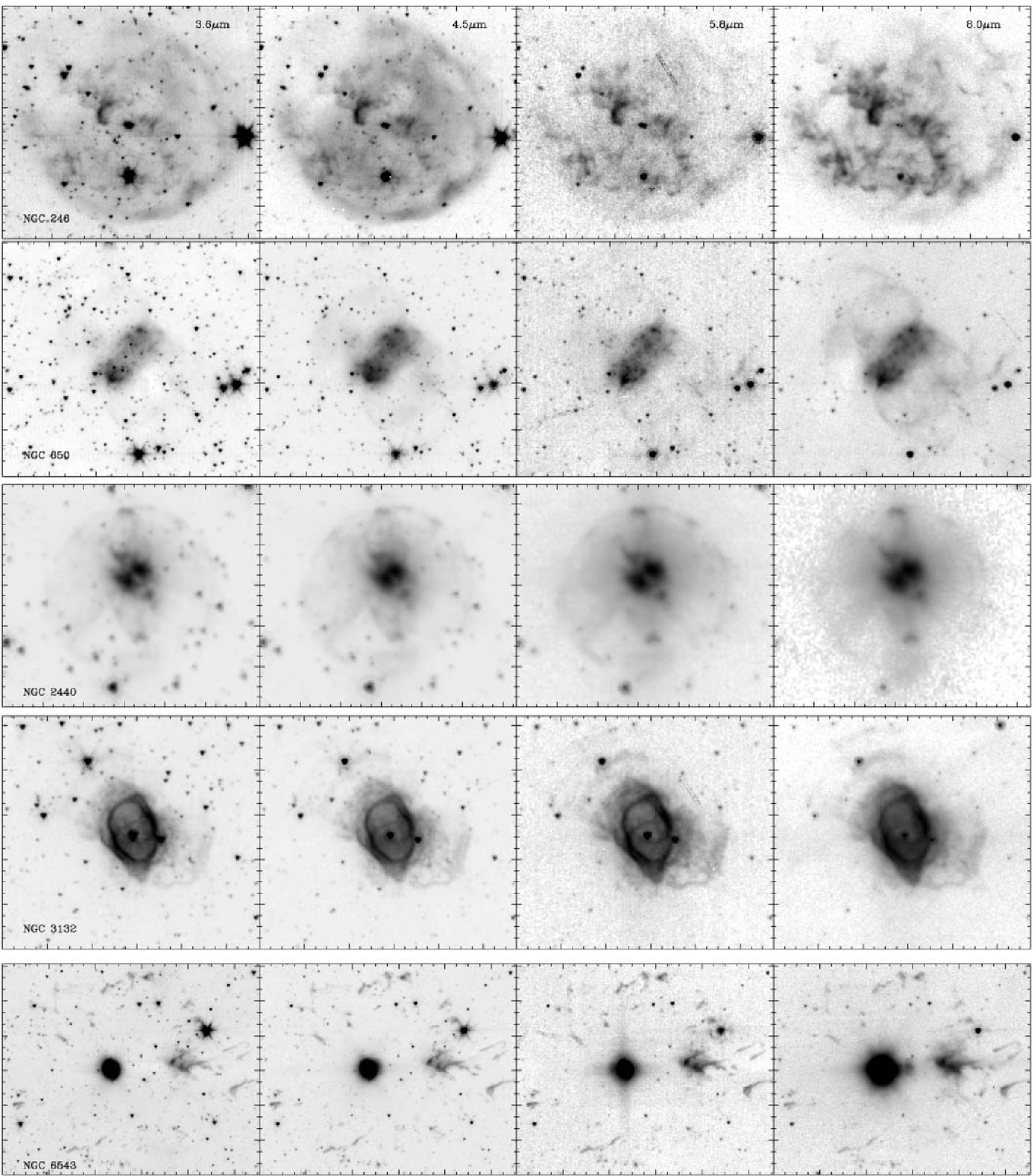}
\caption{(Plate 2): Planetary nebulae images, single IRAC bands.  The images
are arranged with each object in one row, with the four columns being 
the IRAC 3.6, 4.5, 5.8, and 8.0 $\mu$m bands.   The 
orientations are the same as in Figure 1.  The images are scaled logarithmically, with the peak level set to the brightest part of the extended emission in the image.  The image
width for each PN is as follows: NGC 246 - 287\arcsec, NGC 650 - 302\arcsec, NGC 2440 - 116\arcsec, NGC 3132 - 210\arcsec, and NGC 6543 - 406\arcsec. 
}
\end{figure}
\clearpage

\end{document}